\pdfoutput=1
\documentclass[twoside,letterpaper]{soups} 
\usepackage{url}
\usepackage{hyperref,breakurl}

\begin{document}
\conferenceinfo{ %
  Workshop on Inclusive Privacy and
  Security (WIPS): Privacy and Security for Everyone, Anytime,
  Anywhere, held as part of
  Symposium on Usable Privacy and Security
  (SOUPS)}{2015, July 22--24, 2015, Ottawa, Canada.}
\CopyrightYear{2015} %

\title{Accessible Banking: Experiences and Future Directions}
\numberofauthors{2}
\author{
\alignauthor
Bela Gor\\
      \affaddr{Business Disability Forum}\\
      \affaddr{Nutmeg House}\\
      \affaddr{60 Gainsford Street}\\
      \affaddr{London, SE1 2NY, U.K.}\\
      \email{belag@businessdisabilityforum.org.uk}
\alignauthor
David Aspinall\\
       \affaddr{School of Informatics}\\
       \affaddr{Informatics Forum}\\
       \affaddr{10 Crichton Street}\\
       \affaddr{Edinburgh, EH8 9AB, U.K.}\\
       \email{david.aspinall@ed.ac.uk}
}

\date{May 2015}

\maketitle

\section{Modern banking demographics}

There are more than 11 million disabled people in the UK, almost 20\%
of the overall population~\cite{BDF}.  This includes 1.87m people with
sight loss and 0.8m who are severely or profoundly
deaf~\cite{RNIB,ActionHearingLoss}.
Both figures are projected to increase significantly with the ageing
population.  In general, the proportion of people with disabilities is
increasing both because of ageing and because an increasing number of
children born with disabilities are surviving, thanks to medical
advances.

It is sometimes imagined %
that older and disabled people seldom use the
Internet or other modern technologies, but this is unfounded.  The UK
communications regulator Ofcom conducts market surveys of
media literacy: the most recent report~\cite{OfcomMediaUse} found
that %
in 2013, 83\% of adults regularly went online.  %
While nearly all 16-34s are online (98\%), the proportion
of over 65s lags but is increasing quickly (42\% in 2013 vs.\@ 33\% in 2012).
Their method of access varies, for modern devices:

\begin{itemize}
\item \textbf{Tablets} have seen a doubling of usage overall in a year
  (16\% in 2012 to 30\% in 2013).
   Usage in the age group 35-64 has doubled, while use by 65-74s has trebled
   in the same time (from 5\% to 17\%).
 \item \textbf{Smartphones} were used by 62\% of the UK population
   in 2013, %
   with those aged 65-74 are almost twice as likely to use a
   smartphone then compared to 2012 (20\% vs.\@ 12\%).
\end{itemize}
(These figures appear comparable to those for the US, although 
adoption rates are generally higher there~\cite{PewOlderAdults}).  %
Futhermore, offline security technologies are already widely used by 
the disabled demographic, as retail service provision has become
increasingly multi-channel.

\medskip

The first author is Legal Director at Business Disability Forum (BDF),
a UK not-for-profit membership organisation which advises businesses
on regulatory requirements and best practice for disabled customers
and employees.  Sponsoring members include over 400 public and private
sector businesses across numerous sectors.
Perhaps unsurprisingly, the banking sector is the one which
raises most issues concerning security technologies for disabled and
older people.

An industry group survey recently published by the British Banking
Association~\cite{BBA}
found that nearly 2.3 million people aged 70 and over %
are now using Internet banking; 0.6m of these people are 80~or older.
The numbers are still small as percentages of the overall population
(e.g., fewer than 1\% of banking app downloads were by customers aged
70 and over), but banks are paying attention for several reasons.
First, the absolute numbers are significant and the study shows that
there is a faster growth rates for digital services by customers in
their 70s and 80s than for younger generations as the market
gets saturated.  Second, banks know
that the older generations tend to have more wealth, it is worth
competing for them.  Finally,
treating older and disabled people better brings potential 
advantage in reputation.

Despite these motivations, current technology solutions for banking
continue to raise a range of (sometimes familiar) %
usability problems for end users.

\section{Common problems}

In this section we pick out some of the most common security-related
problems which are raised by disabled and older people accessing
modern banking services.  The points are gathered from long experience
in BDF helping businesses manage best practice and their customer
complaints, and supported by quotations published in a recent market
research report \emph{Missing Out} compiled by Really Useful Stuff (a
reseller for assistive products), who surveyed 350 disabled
people on their experiences with consumer banks, among other
retailers~\cite{MissingOut}.

\subsection{Unreasonable security-usability tradeoffs}

On-screen keyboards (and pull-down) menus were introduced into online
banking authentication to address security concerns over keystroke
logging malware.  These are perhaps the most common barrier for people
who use assistive technology, since they are very difficult to use
without a mouse and impossible with screen readers, suck and blow
devices or speech recognition devices.

Telephone banking is another channel, but it also enforces strict
security controls:
\begin{quote}
  Sometimes I phone up for support from a bank, I can answer all the
  security questions but if I say ``I now need to pass you to my
  husband as he can read the screen to you'' my bank's staff refuse to
  talk to him, even though I've just gone through the security
  questions, proved it's me and explained that I'm blind.~\cite{MissingOut}
\end{quote}
The lack of accessibility and usability of security systems can lead to
unanticipated security risks as people try to ``work around'' the
security systems.  There is anecdotal evidence that disabled people pass
on password, PIN and answers to security questions to family members,
friends and in some instances to carers who pretend to be the customer.
This leaves them open to the risk of fraud --- particularly in the case
of carers who may be employed on a temporary basis by the disabled
person.  If the bank discovers that the customer has divulged security
information to a third party it will not reimburse a victim of fraud
because the customer has breached the terms of their contract with the
bank.  So far, banks have back-pedalled rapidly in cases where a
disabled person is concerned. %

\subsection{Accessibility added then removed}

It is often reported that systems that were originally designed to be
accessible and inclusive are not maintained; subsequent revisions or
updates make the accessible inaccessible:
\begin{quote}
  I am totally blind and my bank's
  web site used to be excellent, both for business and personal accounts,
  but then they seemed to lose their grip on web accessibility and now I
  can hardly use their site at all.~\cite{MissingOut}
\end{quote}

\begin{quote}
  My online banking became inaccessible overnight -- they must have
  changed something because my screen reader just couldn't access my
  account information.~\cite{MissingOut}
\end{quote}
The problem is that the updates are not tested properly and are being
made without the user's control.  We know that users in general are wary of
manual software updates because of these kind of
issues~\cite{DBLP:conf/chi/VanieaRW14}; updates of websites and mobile
apps are largely forced on users automatically.

\subsection{Access needs ignored despite regulation}

Different access needs are not always considered when new technology,
systems and processes are introduced, even though there is a legal
requirement in the UK to anticipate the needs of disabled customers.
The Equality Act 2010 requires all providers of goods, services, and
public functions to make ``reasonable adjustments'' for disabled
people, removing barriers to access.  The legal duty is two-fold,
applying in general and for individuals.  Organisations should expect
disabled customers and anticipate their needs: buildings, telphones,
websites, apps, etc, must be user tested to be as accessible and
usable as possible, covering a wide range of disabilities.  If,
despite this, %
a service remains inaccessible for an individual, the organisation
must make specific reasonable adjustments for that person.  Case law,
so far only in the physical domain,
has made it clear that service providers have a legal duty to enable
disabled people to enjoy a service that is the same or as close as
possible to that enjoyed by people without their disability,
encouraging inclusive methods rather than alternative or ``special''
methods of access.
Older banking technologies such as bank credit and debit cards and ATMs
have long caused problems.
ATMs are often too high or set too far back to be used by wheelchair
users and until recently were almost completely inaccessible to people
with visual impairments. Again, security risks arise because
disabled people hand over cards and PINs to third parties to obtain
cash for them.  

Recent Two-Factor~Authentication (2FA) devices and apps 
are problematic for many users: see \cite{KrolEtAl15} for
an insightful user study and evidence of widespread frustration.   
Some 2FA devices are designed to be small enough to fit
into a purse or dangle from a key ring.
This very design means that the push buttons and the numbers and
letters on the screen are too small for many people to either see or
push.  One bank reported an army veteran who had lost an arm in combat
complaining that he couldn't use the 2FA device with one hand.  If he
placed it on a table or desk, it moved around too much for him to be
able to push the buttons. His improvised solution was to tape it to
his desk thereby negating the value of it being small and portable!

\subsection{Training and awareness}

In some instances it isn't the technology that is the barrier but the
lack of training and awareness of staff on what is and isn't possible.
Chip and PIN devices are an example of technology that, as banks and
retailers have realised for some time, is difficult for some disabled
people to use.  To improve accessibility, Chip and PIN machines are
designed to move, but when this doesn't help a person, banks will
issue Chip and signature cards (and where necessary a signature stamp).
Disabled people, however, frequently report that staff in retail
outfits are unaware that their business accepts Chip and Signature or
signature stamps:
\begin{quote}
  I don't use a PIN -- I use Chip and Signature. At {[}name
  withheld{]} checkout, I explained this to the assistant but she
  tried to swipe it. When I explained, she said ``I know what I'm
  doing'' but she didn't. The member of staff argued with me, the
  customer. She seemed totally untrained.~\cite{MissingOut}
\end{quote}
\begin{quote}
  When I asked a shop assistant to pass me the key pad, so I can enter
  my PIN, they said ``it doesn't move!''. I had to ask another person
  in the queue to pull the keypad out of its holder and pass it to
  me.  The shop assistant said ``I didn't know it did that''.~\cite{MissingOut}
\end{quote}
Such experiences lead many disabled people to shop online (and this
recommendation has even been used as an attempted reasonable
adjustment by one company~\cite{RBS-Allen}). 
But that leads us back to inaccessible websites and security
obstacles such as CAPTCHAs which can be difficult for people with a
range of disabilities.  CAPTCHAs often now have accessible
``alternatives'' but this goes against the ideal of 
inclusive design.

\section{New solutions}

The banking industry in the UK is very aware of these accessibility
issues and have taken steps, particularly recently, to address them.
Encouragingly, some of the new solutions seem to be admitting
altering tradeoffs between security and usability.

\textbf{User-friendly cards.}  Special brightly coloured ``high
visibility'' credit and debit cards are available that have notches
and arrows to show which way they should be inserted into ATM or Chip
\& PIN readers.
It is also possible to personalise some of these cards. This enables
people who are partially sighted to identify the card they wish to use
and use it more easily.  Arguably notches and arrows would benefit all
users as a standard inclusive design: confusion over card orientation
is all too common.

\textbf{Sign Video.} This service
allows sign language users to connect to telephony agents via an
accepted third party interpreter using a PC or tablet screen.  It
introduces a new security threat, perhaps, but should prevent
experiences like this deaf customer's:
\begin{quote}
  I prefer to communicate through a BSL [British Sign Language]
  interpreter but this is not always offered. At one meeting with my
  bank, I had to get a member of staff to phone the call centre, only
  for them to mishear the word ``deaf'' and close my account thinking I
  was dead! It ended up with letters to my executor. Trying to
  convince them I was alive took weeks.~\cite{MissingOut}
\end{quote}

\textbf{Talking ATMs.}  Trials by Barclays~\cite{Barclays} of
the first talking ATMs in the UK have gained positive responses:
\begin{quote}
  I've not withdrawn money from an ATM independently for 15 years! The
  talking ATM was a life changing experience.~\cite{MissingOut}
\end{quote}

\textbf{Voice Biometrics.} Several security companies are developing
voice biometric solutions, which some banks are investigating.  The
aim is to avoid the need for customers to remember passwords, PINS and
answers to security questions, particularly benefiting people with
cognitive impairments such as dementia and dyslexia.  Again, this kind
of solution may be a more acceptable tradeoff between security and
usability, accepting potential risks of spoofing~\cite{Wu2015130}.

\textbf{Discreet Beaconing.}  Finally, another innovation being
trialed by Barclays~\cite{Barclays} provides a Bluetooth-enabled
smartphone app that shares disability and identity information
(specified by the user) with customer service staff, in a discreet way
(e.g., a message on the till screen).  A customer might indicate that
she has a hearing impairment but does better when someone speaks close
to her right ear.  Another customer might forewarn that he has
dyslexia and so uses a signature rather than PIN number.  This
certainly raises privacy and security concerns (broadcasting ones
disability generally sounds inadvisable), but addresses a common
frustration.  Reportedly the initiative was started after a customer
in Sheffield suggested that services could be improved for disabled
people, especially by reducing the need for customers to have to
explain their accessibility needs every time they enter the branch.

\section{Future Directions}

Although some UK organisations, notably the large banks, are doing
more to make their services accessible to disabled and older people
there is a long way to go.
Disabled
people still find themselves excluded by access methods,
 and in particular, by security systems.
One of the main reasons for this is that
front line and back office staff are ill-trained in
accessibility.  Front line staff often lack awareness of the
accessibility features in common technology.
Back office staff can render accessible systems inaccessible
by applying ``improvements'' and upgrades without due care. 

Innovations are welcome, but need proper user studies among varied
demographics before mass deployment.  Future research areas 
include investigating whether accessibility can be built-in
robustly, so it cannot be overridden or defeated by humans later.
This may amount to broader use of standards and
(ideally automated) testing for compliance, along with better
frameworks.  For example, web design frameworks should better ensure
universal inclusive design
 (e.g., extending ``responsive'' elements for accessible
elements, alternative input methods).
We also need ways of training many front-line
staff in the accessibility features of technology and how to use them,
in an empathetic way.
Ultimately, the UK law exists to protect disabled people.  But legal
processes are difficult and expensive; current changes are
driven more by commercial and reputational factors.

\bibliographystyle{abbrv}
\bibliography{accessiblebanking} 
\end{document}